\documentclass[journal,letterpaper]{IEEEtran}
\usepackage{amsmath}
\usepackage{amsthm}
\usepackage{amsfonts}
\usepackage{amssymb}
\usepackage{makeidx}
\usepackage{cite}
\usepackage{graphicx,bigints}
\usepackage{mathtools}
\usepackage{cases}
\usepackage{color}
\usepackage{hyperref}
\usepackage{bbm}
\usepackage[normalem]{ulem}
\usepackage{braket}
\usepackage{textcomp,gensymb}

\usepackage{enumitem}
\usepackage{datetime}

\usepackage{epstopdf}
\usepackage{xcolor}
\usepackage{lipsum}

\usepackage{multirow}

\usepackage[makeroom]{cancel}

\makeatletter
\DeclareFontFamily{U}{tipa}{}
\DeclareFontShape{U}{tipa}{m}{n}{<->tipa10}{}
\newcommand{\arc@char}{{\usefont{U}{tipa}{m}{n}\symbol{62}}}%

\newcommand{\arc}[1]{\mathpalette\arc@arc{#1}}

\newcommand{\arc@arc}[2]{%
	\sbox0{$\m@th#1#2$}%
	\vbox{
		\hbox{\resizebox{\wd0}{\height}{\arc@char}}
		\nointerlineskip
		\box0
	}%
}
\makeatother

\newtheorem{theorem}{Theorem}

\newtheorem{lemma}{Lemma}

\newtheorem{example}{Example}

\newtheorem{remark}{Remark}

\DeclareMathOperator*{\argmax}{arg\,max}

\DeclareMathOperator{\cO}{\mathcal{O}}

\DeclareMathOperator{\cL}{\mathcal{L}}

\DeclareMathOperator{\SINR}{\textrm{SINR}}

\DeclareMathOperator{\bP}{\mathbf{P}}
\DeclareMathOperator{\bS}{\mathbb{S}}

\DeclareMathOperator{\bE}{\mathbf{E}}

\newcommand*\diff{\mathop{}\!\mathrm{d}}

\newcommand*\nnb{\nonumber}

\newcommand{\overbar}[1]{\mkern 1.5mu\overline{\mkern-1.5mu#1\mkern-1.5mu}\mkern 1.5mu}

\definecolor{sandy}{HTML}{E6E2AF}
\definecolor{stone}{HTML}{A7A37E}
\definecolor{beach}{HTML}{EFECCA}
\definecolor{ocean}{HTML}{046380}
\definecolor{diver}{HTML}{002F2F}

\definecolor{Firenze1}{HTML}{468966}
\definecolor{Firenze2}{HTML}{FFF0A5}
\definecolor{Firenze3}{HTML}{FFB03B}
\definecolor{Firenze4}{HTML}{B64926}
\definecolor{Firenze5}{HTML}{8E2800}
\definecolor{mediumpersianblue}{rgb}{0.0, 0.4, 0.65}
\definecolor{hongik}{HTML}{004498}
\definecolor{cobalt}{rgb}{0.0, 0.28, 0.67}
\definecolor{burntorange}{rgb}{0.8, 0.33, 0.0}

\definecolor{ultramarineblue}{rgb}{0.25, 0.4, 0.96}

\title{Stochastic Geometry and Dynamical System Analysis of Walker Satellite Constellations}

\author{Chang-Sik Choi
  	 and Francois Baccelli
	\IEEEcompsocitemizethanks{\IEEEcompsocthanksitem{Chang-Sik Choi is with School of Electrical Engineering, KAIST, South Korea.  (email: changsik@kaist.ac.kr).
			Francois Baccelli is with Telecom Paris and Inria Paris (email: francois.baccelli@ens.fr).
			The work of C. Choi was supported by NRF RS-2024-00334240.
			The work of F. Baccelli was supported by the European
		        Research Council NEMO project (grant ERC 788851),
		        the Horizon Europe INSTINCT project (grant SNS 101139161),
		        the France 2030 projects PEPR réseaux du Futur project
		        (grant ANR-22-PEFT-0010), and by 5G NTN mmWave (BPIFrance).
			This joint work was also supported by a South Korea--France Hubert Curien grant.}
}
\IEEEcompsocitemizethanks{\IEEEcompsocthanksitem{The authors thank the reviewers as well as Patrick Thiran of EPFL and Rudrapatna Ramakanth of MIT for their comments which helped improving this paper.  R. Ramakanth asked in particular whether any satellites collide in the Walker model introduced in this paper. The answer to this question is given in the appendix of the present paper.}}
}
\begin{document}
	\maketitle

\begin{abstract}
In practice, low Earth orbit (LEO) and medium Earth orbit (MEO) satellite networks consist of multiple orbits which are populated with many satellites. A widely used spatial architecture for LEO or MEO satellites is the Walker constellation, where the longitudes of orbits are evenly spaced and the satellites are equally spaced along the orbits. In this paper, we develop a stochastic geometry model for the Walker constellations. This proposed model enables an analysis based on dynamical system theory, which allows one to address essential structural properties such as periodicity and ergodicity. It also enables a stochastic geometry analysis under which we derive the performance of downlink communications of a typical user at a given latitude, as a function of the key constellation parameters.
\end{abstract}


\section{Introduction}
\subsection{Motivation and Related Work}
\IEEEPARstart{L}{EO} and MEO satellite networks are designed to support various applications, including data communications, data sensing and harvesting, and Internet routing. Geometrically, LEO and MEO satellites are located on specific orbits, which determine not only their relative positions but also their motions. Practical satellite networks are primarily designed to maximize coverage based on two characteristic geometric features: (i) orbit longitudes are distributed evenly over the equator, and (ii) satellites are placed evenly within each orbit \cite{FCCSTARLINK}. This configuration is known as the Walker constellation. From the perspective of a fixed observer on Earth, such a Walker constellation is often presumed to lead to periodic \emph{ephemerides}, given the regular distribution of both orbits and satellites. But is it truly the case?

Recently, stochastic geometry has been widely used to model satellite networks by mathematically characterizing the spatial distribution of satellites. For instance, binomial and Poisson point processes have been employed to represent satellites as uniform random points on a sphere \cite{9079921,9177073}. However, these models fail to capture the orbital structure of practical LEO and MEO satellite networks, where satellites are clustered along orbits. To address this, Cox point processes were proposed to jointly model both the distribution of orbits and the satellites positioned along them \cite{10410220,10436110,10557592,10703111,10771991}. Although Cox point processes account for orbit-based clustering, they typically assume isotropic configurations, similar to binomial or Poisson models. Consequently, essential dynamical system properties of non-isotropic Walker constellations---such as periodicity and ergodicity---cannot be rigorously studied within the existing stochastic geometry frameworks. To examine the dynamics of Walker constellations, we develop dedicated stochastic geometry models for them and then analyze their statistical properties.

\subsection{Theoretical Contributions}

First, we develop a stochastic geometry framework to represent the locations of satellites in a Walker constellation. The proposed framework features a non-isotropic network architecture whose geometry is determined by network parameters including the orbit inclination $\phi$, the number of orbits $N_o$, the number of satellites per orbit $N_s$, and the user latitude $l_u$.

Using this framework, we derive essential statistical metrics crucial to the performance of the satellite communications. First, we show that the proposed model has a distribution which is invariant with time. Then, we identify the exact analytic condition under which the proposed model is periodic or ergodic. The ergodicity means that the expectation of a snap shots of network performance experienced by the typical user at a given latitude \emph{is equal to} the empirical average of network performance of any user at the same latitude over a very long time. Then, assuming the typical user associates with its nearest satellite, we derive the distribution of the distance from the typical user to its serving satellite as a function of orbit inclination, the number of orbits, the number of satellites, and the user latitude, namely $\phi$, \(N_o\), \(N_s\), and $l_u$, respectively. Moreover, in contrasts to existing stochastic geometry models, the proposed model inherently exhibits an upper-bounded association distance, which, under appropriate conditions, lead to a lower bound on the rates of communications from satellites to users on Earth. Assuming a simple antenna gain model for satellites, we derive formulas for the Laplace transform of the total interference power and its expectation, respectively. They are crucial for understanding the amount of interference seen by a typical user at any latitude and therefore can be used to design interference mitigation techniques. Finally, we evaluate the coverage probability, namely the signal to interference-plus-noise ratio (SINR) distribution of the link from the association satellite to the typical user at a given latitude.

\section{System Model}
\begin{figure}
	\centering
	\includegraphics[width=.7\linewidth]{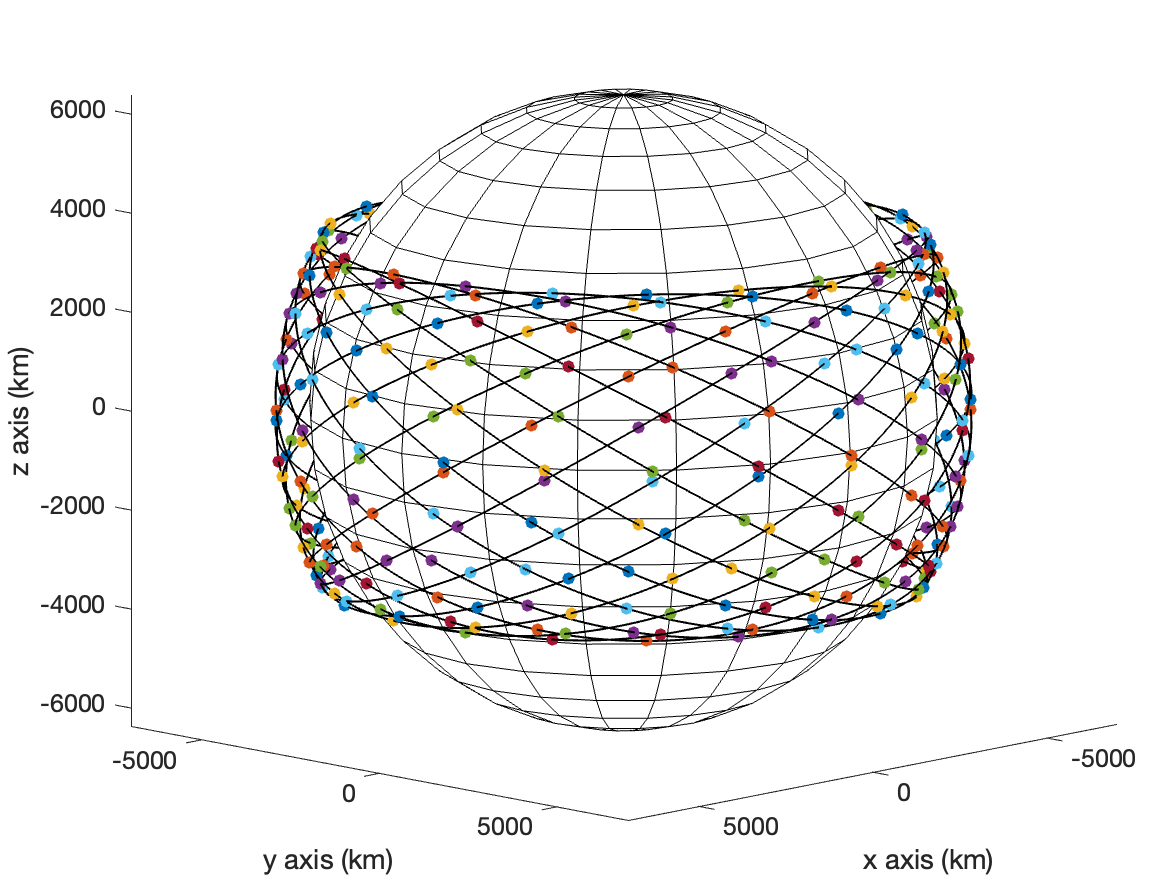}
	\caption{The proposed newtork with orbits and satellites regularly distributed.}
	\label{fig:spatialmodel}
\end{figure}

\subsection{Spatial Model and Signal Propagation}

In this paper, the three-dimensional space is equipped with the orthonormal basis $(e_x,e_y,e_z)$ with the origin at the center of Earth. The reference space is rotating with the Earth's natural spin, so that a fixed location on Earth has constant coordinates. The $xy$-plane is the equatorial plane, the $x$-axis is the longitude reference axis and the $z$-axis is the North Pole axis.

The network users are randomly distributed at fixed locations on the surface of Earth of radius $e=6370$ km. We assume that all satellite orbits are circular. They do not rotate with the Earth. The satellites on the orbits rotate around the Earth, all in the same direction, with a constant speed.

At a given time, the ascending point of an orbit is the intersection point of the orbit with the reference plane. This is the point where satellites cross the Equator from South to North. The longitude of an orbit is defined as the angle from the $x$-axis to the ascending point, measured on the equatorial plane. The inclination of an orbit is defined as the angle that the orbital plane makes with the equatorial plane at the ascending point. The inclination does not change with time.  The phase of a satellite at a given time is defined as the angle between the satellite and the ascending point at that time. This angle is measured in the orbital plane of the satellite.

Let $N_{o}$ be the number of orbits and $N_s$ be the number of satellites on each orbit. We assume that the longitudes of the orbits are evenly distributed, namely those $N_o$ ascending points are distributed evenly on the  $[0,2\pi]$ interval. In addition, we assume that, at time zero, a uniformly distributed random offset $\bar{\theta} \sim \text{Uniform}(0,2\pi/N_o)$ is added to the ascending points. Consequently, at time zero, the longitude of the $i$-th orbit ascending point is given by
\begin{equation}
	\theta_i=2\pi i  /N_o + \bar{\theta}  \mod 2\pi, \label{eq:theta_i}
\end{equation}
for $i=1,\ldots,N_{o}$. Note that for all $i$, the same offset $\bar{\theta} $ is added.
This random shift of periodic points leads to longitudinal rotation invariance, or equivalently, invariance of orbit
distribution w.r.t. the Earth's spin. The inclinations of orbits are assumed to be all equal to $\phi$ and
the orbit process at time 0 is denoted by $\cO = \bigcup_{i=1}^{N_o} l_{i} = \bigcup_{i=1}^{N_o}  l(\theta_{i},\phi)$ where $l_i(\theta_i,\phi)$ is the $i$-th orbit, with longitude angle $\theta_i$ and inclination angle $\phi.$

Conditionally on these orbits, satellites are distributed regularly on each orbit at time 0,
namely the phases of the $N_s$ satellites on each orbit are evenly distributed over $[0,2\pi].$
Let $\bar{\omega} \sim \text{Uniform}(0,2\pi/N_s)$
be an independent random variable. The phase of the $j$-th satellite on the $i$-th orbit at time zero is
\begin{equation}
	\omega_j=  2\pi j/N_s + \bar{\omega} \mod 2\pi\label{eq:omega_j},
\end{equation}
for $j=1,\ldots,N_s$. In Eq. \eqref{eq:omega_j}, the same random variable $\bar{\omega}$ is added to the phases of \emph{all} satellites. 
Collectively, the satellite point process at time zero is given by
\begin{equation}
	\Psi = \sum_{i=1}^{N_o} {\psi_i}= \sum_{i=1}^{N_o}  \sum_{j=1}^{N_s}\delta_{X_{i,j}} \label{eq:psi},
\end{equation}
where $\psi_i$ is the satellite point process on the $i$-th orbit $l_i(\theta_i,\phi)$ and $\delta_{X} $
is the Dirac measure at $X$, and $X_{i,j}$ is the $j$-th satellite with its phase $\omega_j$ on the orbit $l_i(\theta_i,\phi)$.

For $i=1,...,N_o$, $j=1,...,N_s,$ the $(x,y,z)$-coordinates of the satellite at $X_{i,j}$ at time zero are given by
\begin{align}
	x&=r\sqrt{\cos^2(\omega_j)+\sin^2(\omega_j)\cos^2(\phi)}\cos(\hat{\theta}_{j}+\theta_{i}),\label{vecx}\\
	y&=r\sqrt{\cos^2(\omega_j)+\sin^2(\omega_j)\cos^2(\phi)}\sin(\hat{\theta}_{j}+\theta_{i}),\label{vecy}\\
	z&=r\sin(\omega_j)\sin(\phi)\label{vecz},
\end{align}
respectively with $\hat{\theta}_{j}= \mathrm{atan2}(\sin(\omega_j)\cos(\phi),\cos(\omega_j))$ where $\mathrm{atan2}$ is a four quadrant arctan function\cite{press2007numerical}. The proposed model is a stochastic geometry formulation of the well-known Walker-delta constellation \cite{walker1984satellite},
especially when the relative phases between satellites are set to be zero,
which can be denoted by a Walker $(\phi: N_oN_s/N_o/0\degree).$

Fig. \ref{fig:spatialmodel} depicts the proposed stochastic geometry model with $N_o = 20$,  $N_s=20,$ $\phi=33\degree,$
and $r=7000$ km. 
The proposed model for the satellite network contrasts with the existing stochastic geometry models based on
binomial point processes \cite{9079921} or an isotropic Cox point process \cite{10410220,10436110,10557592,10703111,10771991},
where both models have distributions invariant w.r.t. all rotations. In those models, the network users anywhere on Earth
see the same satellite distribution, resulting in an average performance metric which is the same for all users on Earth, regardless of the users' latitudes. In contrast, in the proposed model, users at different latitudes experience different satellite distributions. Consequently, network performance metrics vary depending on the user's latitude. This aspect will be explained further.

In this paper, the typical user's longitude is assumed to be $0\degree$ and the typical user is at $\vec{u} = \left(e \cos(l_u), \ 0, \ e\sin(l_u)\right)$, where $l_u$  denotes the user latitude in the range: $ -\pi/2 \leq l_u\leq \pi/2.$

For performance analysis, all satellites are assumed to use the same wireless resource and network users on Earth are associated with their nearest satellites. As in \cite[6.1.3]{38821}, we assume the received signal power is given by a power-law path loss, multiplied by a random fading and by the antenna gain of the corresponding propagation channel. For the channel of distance $d>1$ (meter), the received signal power is $	pG(d)Hd^{-\alpha} $ with $p$ the average isotropic received signal power at $d=1$, $G(d)$ the aggregate antenna gain, and $H$ the random fading over the channel.
For $G(d),$ we assume that satellite antennas provide additional gains of $g_t$ and the receive antenna provides an isotropic gain of $g_r$. Inspired by the practical guideline in \cite{38821}, we assume the antenna gain over the channel of length $d$ is given by $G(d)=g_tg_r$ if $d\leq d_g;$ and $G(d) =g_r  $ if $  d>d_g$
where $d_g$ is the cutoff distance\footnote{This follows from \cite{38821} where ground users within a certain angle have an antenna gain $g_t$. }. The cumulative CDF (CCDF) of $H$ is denoted by $\bar{F}_H(x)$ and we assume $\bE[H]=1$.



In this paper, we first examine the dynamical aspect of the proposed stochastic geometry network model. We then derive the distance distribution from the typical user to its nearest satellite, the Laplace transform of signal-plus-interference power of the typical user, and finally the coverage probability of the typical user, defined as the probability that the SINR of the typical user is greater than a given threshold $\tau.$

\section{Dynamical System Analysis}

Let $S$ be a measurable space.
A dynamical system on $S$ is a measurable flow $\{R_t\}_{t\in \mathbb R^+}$ on $S$,
namely a family of deterministic maps $R_t$ from $S$ to itself,
such that (i) the map $(t,x)\to R_t(x)$ is measurable from
$\mathbb R^+\times S$ to $S$, and (ii) $R_t\circ R_s=R_{t+s}$,
for all $s,t\in \mathbb R^+$.
The state of the dynamical system at time $t$ is
$R_t(x)$, with $x$ the initial state, namely $R_0(x)=x$.

Moreover, it is said to admit an {\em invariant measure} if there exists a probability measure $\mathcal P$ on $S$ such that
if $x$ is chosen at random with the distribution $\mathcal P$, then $R_t(x)$
has $\mathcal P$ for distribution for all $t\ge 0$.

In case $S$ is a metric space equipped with its Borel $\sigma$-algebra and the map $(t,x)\to R_t(x)$ is continuous,
the flow is said to be \emph{minimal} if, the trajectory $R_t(x)$ is dense in $S$ for all $x\in S$.

Let $\{R_t\}$ be a measurable flow and
let $h:S \to \mathbb{R}$ be a measurable function.
The time average of $h(R)$ associated with the initial condition $x$ is defined as
$\lim_{T\to \infty} \frac 1 T \int_0^T h(R_t(x)) \mathrm{d} t$,
when the limit exists.
If $\mathcal P$ is an invariant probability measure for $R$, then,
for all measurable functions $h:S \to \mathbb{R}$ such that
$\int |h(s)| {\mathcal P}(\mathrm{d} s)<\infty$, the time average of $h(R)$
exists for
$\mathcal P$ almost-all $x$ (this is Birkhoff's theorem \cite{Katok}). The dynamical system $R$ is said to be
{\em ergodic} w.r.t. $\mathcal P$ if, for all such $h$,
and for $\mathcal P$ almost-all $x$,
\begin{equation}
\lim_{T\to \infty} \frac 1 T \int_0^T h(R_t(x)) \mathrm{d} t =
	\int h(s) {\mathcal P}(\mathrm{d} s),\label{eq:ergodic0}
\end{equation}
i.e., the long term time average of $h(R)$ is the
mean of $h(R)$ w.r.t. the invariant distribution $\mathcal P$.

\begin{example}
Consider the measurable flow $R^{\alpha}_t(x)$ on $S=[0,1]$,
where $R_t^{\alpha}(x) = x + \alpha t \mod 1 $.
Here, $x$ is the initial condition, $\alpha$ is the velocity, and $R_t^\alpha(x)$ is the state at time $t$.
Note that $S$ is a compact metric space and that this flow is continuous.
The uniform measure on $[0,1]$, $\mathcal P$,
is an invariant probability measure of $R^{\alpha}$.
If $\alpha$ is rational, the dynamic is periodic and is hence neither minimal nor ergodic w.r.t. $\mathcal P$.
If $\alpha$ is irrational, then the dynamical system is minimal and ergodic w.r.t. $\mathcal P $.
\end{example}

In the proposed Walker satellite model,
let $ \bar{\theta}_t$ denote the longitude of the first orbit (that with the smallest longitude) at time $t$ and
$\bar{\omega}_t$ the angle of the first satellite on this orbit (that with the smallest phase) at time $t$.
Let $\bar{\theta}$ and $\bar{\omega}$ denote the initial condition of these state variables. Then,
\begin{align}
	\bar{\theta}_t(\bar\theta) = \bar{\theta} - \bar v_{{\theta}} t  \hspace{-2mm} \mod \frac{2\pi}{N_0} \text{ and }
	\bar{\omega}_t(\bar\omega) = \bar {\omega}+ \bar v_{{\omega}} t  \hspace{-2mm}  \mod \frac{2\pi}{N_s},\nnb
\end{align}
where $\bar v_{\theta} $ and $\bar v_{\omega}$ denote the
angular speed of Earth-spin and the angular speed of satellite rotation, respectively.
Our key measurable flow is $R_t(x)= (\bar{\theta}_t(\bar\theta), \bar{\omega}_t(\bar\omega))$,
with state space $S=[0,\frac{2\pi}{N_o})\times[0,\frac{2\pi}{N_s})$ and with initial condition
$x=(\bar {\theta},\bar {\omega})$.

We now describe the main ideas developed in the forthcoming subsections.
Let $\Psi_t$ denote the satellite point process at time $t$, namely the point process $\Psi$ in
Eq. \eqref{eq:psi}, with $\bar \theta$ replaced by $ \bar\theta_t(\bar\theta)$
and $\bar \omega$ replaced by  $\bar\omega_t(\bar\omega)$, respectively.
Since this point process is described in the referential space that rotates with Earth,
in this referential, a static Earth user is a point with static coordinates.
Hence, $\Psi_t$ provides a complete description of the {\em ephemerides} of the Walker constellation.
The space $\mathcal{N}_r$ of counting measures, namely the sphere of radius $r$
where the satellites move, is a measurable space.

Here are two important observations: (i) $\{\Psi_t\}$ is a measurable flow on
$\mathcal{N}_r$; (ii) for all $t\ge 0$,
$\Psi_t$ is a deterministic measurable functional $\Xi$ of $R_t$.
The map $\Xi$ is measurable from $S$ to $\mathcal{N}_r$.
Therefore, if $\{R_t\}$ has an invariant probability measure $\mathcal P$ on $S$, then $\{\Psi_t\}$
has an invariant probability measure as well,
which is the push-forward of $\mathcal P$ by $\Xi$ on $\mathcal{N}_r$, denoted by $ {\mathcal P}_{\Psi}$, which is
a probability measure on $\mathcal{N}_r$. Note that
${\mathcal P}_{\Psi}$ is the distribution of $\Psi_0$ when
$(\bar \theta,\bar \omega)$ has the distribution $\mathcal P$. In addition,
if $\{R_t\}$ is ergodic w.r.t. $\mathcal P$, then $\{\Psi_t\}$ is ergodic w.r.t. $ {\mathcal P}_{\Psi}$.
These last properties can also be understood as the {\em measurable conjugacy} of the dynamical systems $\{R_t\}$ and $\{\Psi_t\}$ (see, e.g., Definition 6.8.2 in \cite{frenchauthor}).

\subsection{Invariant Probability Measure}
Let $\overline{\mathcal Q}$ denote
the product of two uniform distributions on the product space $S=[0,\frac{2\pi}{N_o}]\times [0,\frac{2\pi}{N_s}]$.
Let $\overline{\mathcal Q}_{\Psi}$ denote the distribution of $\Psi$ at time zero when
$(\bar \theta,\bar \omega)$ has distribution $\overline{\mathcal Q}$.

\begin{theorem}\label{Theorem:0} The probability measure
	$\overline{\mathcal Q}$, on $S$, is invariant for $\{R_t\}$ and the probability measure
	$\overline{\mathcal Q}_{\Psi}$, on $\mathcal{N}_r$, is invariant for $\{\Psi_t\}$.
\end{theorem}
\begin{IEEEproof}
It is enough to prove the first statement.
Let $\mathbb{T} [0,a]\times [0,b]$ denote the torus on $[0,a]\times [0,b]$, with $a,b$ positive real numbers.
Let $\theta= \bar\theta\frac{N_o}{2\pi}$ and $\omega= \bar\omega\frac{N_s}{2\pi}$.
Let also $\theta_t= \bar\theta_t\frac{N_o}{2\pi}$
and $\omega_t= \bar\omega_t\frac{N_s}{2\pi}$. Denote by $g$ the linear map that transforms
$(\theta_t,\omega_t)$ into $(\bar \theta_t,\bar \omega_t)$. The associated measurable flow is now $\alpha_t ({\theta},{\omega}) = ({\theta}_0 - v_{{\theta}} t   \mod 1, {\omega}_0 + v_{{\omega}} t   \mod 1)$,
where $v_{{\theta}} = \bar v_{{\theta}}\frac{N_o}{2\pi}$ and $v_{{\omega}} = \bar v_{{\omega}}\frac{N_s}{2\pi}$.

Then, $\overline{\mathcal Q}$ on $\bar{\mathbb{T}}([0,{2\pi}/N_o],[0,{2\pi}/N_s])$ is invariant for
$\bar{\alpha}_t (\bar{\theta},\bar{\omega}) =  (\bar{\theta} - \bar v_\theta t   {\mod 2\pi/N_s},
\bar{\omega} + \bar v_{\omega} t   {\mod 2\pi/N_o})$ iff  $\mathcal Q$ on $\mathbb{T}([0,1],[0,1])$,
namely the product of the two uniform distributions on $[0,1]$, is left invariant by $\alpha_t(\theta,\omega)$.
Then, $\overline{\mathcal{Q}}$ is left invariant by $\alpha_t(\theta,\omega)$ for all nonzero $(v_{{\theta}},v_{{\omega}})$ \cite{Katok}.
\end{IEEEproof}

\subsection{Ergodicity and Periodicity}
Below, we denote by $\rho$ the rotation speed ratio ${\bar v_{\theta}}/{\bar v_{\omega}}$.

\begin{theorem}	\label{Theorem:periodic}
	If the ratio $\rho$ is irrational, then the dynamical system $\{R_t\}$ (resp. $\{\Psi_t\}$)
	on $S$ (resp. ${\mathcal N}_r$)
	is minimal and ergodic w.r.t. $\overline {\mathcal Q}$
	(resp. ergodic w.r.t. ${\overline{\mathcal Q}}$).
	If the ratio $\rho$ is rational, then $\{R_t\}$ and $\{\Psi_t\}$ are both periodic.
\end{theorem}
\begin{IEEEproof}
It is enough to prove the statement concerning $R$.
We use the notation in the proof of Theorem \ref{Theorem:0}.
Let $\hat{\rho}$ denote the normalized ratio ${v_\theta}/{v_\omega}$.
If $\hat{\rho}$ is irrational, then the dynamical system $\alpha_t$ on
$\mathbb{T}([0,1],[0,1])$ is minimal and ergodic w.r.t. $\mathcal Q$, namely the product
of two uniform distributions on $[0,1]$ (see e.g. \cite[Section 6.5.2]{frenchauthor} and
\cite[Theorem 2.1]{carrand2022logarithmic}).
If $\hat{\rho}$ is rational  (which is equivalent to having the ratio $\rho $ rational),
then $\alpha_t$ is periodic on $\mathbb{T}([0,1],[0,1])$. Hence, it is not ergodic w.r.t. $\mathcal Q$
(see \cite[Section 6.5.2]{frenchauthor} and \cite[Theorem 2.1]{carrand2022logarithmic}).
\end{IEEEproof}

\begin{remark}
In the irrational case,	the constellation seen for a fixed Earth user is not periodic.
Instead, we have
\begin{equation}
\lim_{T\to \infty} \frac 1 T \int_0^T h(\Psi_t(x)) \mathrm{d} t =
\int h(\psi) {\overline{\mathcal Q}}_{\Psi}(\mathrm{d} \psi),\label{eq:ergodic}
\end{equation}
for all functions $h: {\mathcal N}_r\to \mathbb{R}$ such that
$\int |h(\psi|) {\overline{\mathcal Q}}_{\Psi}(\mathrm{d} \psi)<\infty$,
where the limit holds for $\overline{\mathcal Q}$ almost all $x$.
In other words, any time average is given by the
expectation w.r.t. the time-invariant distribution ${\overline{\mathcal Q}}_{\Psi}$ of $\Psi$.
For instance $h$ could be the SINR offered by the constellation to a user located at $(x,y,z)$ on Earth.
In the rational case, the system is periodic. As a result, the empirically-obtained time average of what is observed by this Earth user
{does not} match averages over the stationary distribution
$\overline{\mathcal{Q}}_{\Psi}$ in general. Since the base point process $\Psi_t$ for the satellite patterns
is periodic in the Earth referential, any process determined by it is also periodic.
For instance, suppose a user is at the fixed position $(x_{1},y_{1},z_{1})$ on Earth. Then,
the satellite pattern seen from this user is periodic too. These periodic patterns observed however
depend on the user location $(x_{1},y_{1},z_{1})$. For another user at $(x_2,y_2,z_2)$, the satellite
patterns seen are also periodic but with a structure possibly different from that of the one located at $(x_{1},y_{1},z_{1})$.
\end{remark}

As shown above, the relative orbital speed of satellites w.r.t. the speed of rotation of
Earth determines whether the constellation is periodic or not. Let us examine this further.
\begin{remark}
	Since the altitude basically determines the satellite speed, the designer can choose an
	altitude making the constellation periodic to fixed Earth observers.
	However, an unaware designer who would pick a real number, say between $350$ km to $1500$ km,
	would get an irrational ratio for almost all choices w.r.t. the Lebesgue measure on this interval
	since the rational numbers have Lebesgue measure zero in the given interval.
	Therefore, based on Theorems \ref{Theorem:0} and \ref{Theorem:periodic}, one can claim that designers
	unaware of this dynamical system property have ``almost surely" built constellations which are
	aperiodic and ergodic, and where the long time empirical averages converge to the expectation
	over the invariant distribution $\overline {\mathcal{Q}}_\Psi$ studied in the present paper.
\end{remark}


\begin{remark}
Assume that the designer selects the angular parameters $\bar \theta$ and $\bar \omega$ as random variables
sampled independently, with each of them being uniform on the relevant interval. Then, when deconditioning the
position of the satellites of the constellation w.r.t. these two random variables, one gets a distribution for
the satellite positions which is invariant over time, and which is precisely the
distribution ${\overline {\mathcal Q}}_{\Psi}$ of this paper and leads to the specific distributions
that we evaluate below for the metrics of interest.
By the strong law of large numbers, the distributions computed here can hence also be seen as the empirical
averages of the distributions evaluated over a large collection of samples of the constellation
obtained from i.i.d. samples of these two angles. This observation is valid in all cases including the periodic case.
Of course, in the case of ergodicity, the metrics evaluated below are also time averages.
\end{remark}




\section{Performance Analysis}
\subsection{Distance to Nearest Satellite}
Let $\bS$ be the sphere of radius $r.$ For $r-e<d<\sqrt{r^2-e^2},$ we define the spherical cap w.r.t. the typical user $u$ and the distance $d$ as follows: $	\bS_u(d) = \{(x,y,z)\in\bS| \|(x,y,z)- \vec{u} \|\leq d \}$. In particular, $\bS_u(\sqrt{r^2-e^2})$ is the \emph{visible} spherical cap from the typical user and we simply denoted it by $\bS_u.$ Let $D(l_u)$ be the distance from the typical user at latitude $l_u $ to its nearest satellite. We set $D(l_u)=\infty$ if no satellite is visible from the typical user.
\begin{theorem}\label{Theorem:1}
	For $r-e<d<\sqrt{r^2-e^2}$, $\bP(D(l_{u})>d)$, the CCDF of $D(l_u)$ is given by
	\begin{equation}
\zeta\int_{0}^{\frac{2\pi}{N_o}}\int_{0}^{\frac{2\pi}{N_s}}  \!\!\mathbbm{1} \left\{ \frac{r^2+e^2-d^2}{2 r e }> \max \frac{\vec{X}_{i,j} \cdot \vec{u} }{\|\vec{X}_{i,j}\|\|\vec{u}\|} \right\}\diff \omega \diff \theta,\nnb
	\end{equation}
	where $\vec{X}_{i,j} $ are in Eqs. \eqref{vecx}--\eqref{vecz}, and $\zeta = N_oN_s/(4\pi^2).$
\end{theorem}

\begin{IEEEproof}
Let $\mathbbm{1}\{x\}$ be the indicator function taking value one if $x$ occurs and zero otherwise. Conditionally on $\bar{\theta},\bar{\omega},$ the distance from the typical user to its nearest satellite is
\begin{align}
	&\bP(D({l}_{u})>d|\bar{\theta},\bar{\omega})\! =\!\mathbbm{1}\!\!\left\{\!\kappa <\min_{(i,j)}\left\{\!\arccos\left(\frac{\vec{X}_{i,j} \cdot \vec{u} }{\|\vec{X}_{i,j}\|\|\vec{u}\|}\right)\!\right\}\!\right\},\nnb
\end{align}
where $(i,j)\in([N_o],[N_s])$ stands for $i=1,...,N_o$ and $j=1,...,N_s.$ To get this, we use the fact that all points of such orbits are not in the cap if and only if the central angle between the user and the satellite closest to the user is greater than $\kappa, $ where $\kappa=\arccos\left(\frac{e^2+r^2-d^2}{2 r e}\right)$ is the central angle from the user to the rim of the cap $\bS_u(d)$. Then,
\begin{align*}
	&\bP(D(l_u)>d)\nnb\\
	&=\zeta\int_{0}^{\frac{2\pi}{N_o}}\int_{0}^{\frac{2\pi}{N_s}}  \mathbbm{1} \left\{ \cos(\kappa)> \max_{(i,j)} \frac{\vec{X}_{i,j} \cdot \vec{u} }{\|X_{i,j}\|\|u\|} \right\}\diff \bar{\omega} \diff \bar{\theta}\\
	&=\zeta\int_{0}^{\frac{2\pi}{N_o}} \! \int_{0}^{\frac{2\pi}{N_s}}  \!\! \mathbbm{1} \left\{ \frac{r^2+e^2-d^2}{2 r e }> \max_{(i,j)} \frac{\vec{X}_{i,j} \cdot \vec{u} }{\|\vec{X}_{i,j}\|\|\vec{u}\|} \right\} \! \diff \bar{\omega} \diff \bar{\theta},
\end{align*}
where we used the fact that $\cos(x)$ decreases over $[0,\pi]$.
\end{IEEEproof}

Thanks to the deterministic structure of the deployment of satellites, for latitudes $l_{u}$ smaller
than a threshold $L$, for $N_o$ and $N_s$ large enough, there exists a constant $d_c(l)\le \sqrt{r^2-e^2}$ such that
$\bP[r-e\leq D(l_{u})< d_c(l_u)]=1$. 
For instance, when $N_o=30$ and $N_s=50,$ we have $\bP(r-e\leq D({l_{u}})<d_c(l_{u})) =1$ with $d_c(l)\approxeq 775 $km, e.g., the association distance is for sure less than $775$ km. This property is useful to establish performance lower bounds since it directly links to the maximum path loss of the downlink communications. The existence of this critical distance comes from the fact that proposed network features regularly placed satellites with a rigid structure, in contrast to other models based on binomial, Poisson, or Cox point processes \cite{9079921,9177073,10410220,10436110,10557592,10703111,10771991}. Finding an explicit expression for $d_c(l)$ is left for future work.

\begin{example}
	Suppose $\rho$ is irrational. Then, the dynamical system is ergodic and based on Theorem \ref{Theorem:0}, we have
	\begin{equation}
		\lim_{T\to \infty} \frac 1 T \int_0^T D(l_{u},t) \mathrm{d} t = \mathbb{E}_{\overline{\mathcal{Q}}_{\Psi}} [D(l_{u})], \ {\overline{\mathcal{Q}}_{\Psi}} \text{ a.s.}
	\end{equation}
	where the left-hand side is the time average of the association distance for a user and the right-hand side is its expectation w.r.t. the invariant distribution, that was obtained in Theorem \ref{Theorem:1}.
\end{example}



Now, we derive the total interference, namely the amount of signal-plus-interference powers seen by the typical user.
\begin{theorem}\label{Theorem:2}
	The Laplace transform of the total interference of the typical user is given by
	\begin{align}
		\cL_T(s)=k\int_{0}^{\frac{2\pi}{N_o}}\int_{0}^{\frac{2\pi}{N_s}} 	e^{\left(\sum {c_{i,j}}\log\left(\cL_H\left(\frac{s		p G_{i,j}}{\|X_{i,j}-u\|^{\alpha}}\right)\right) \right)}\diff \bar{\omega} \diff \bar{\theta},\nnb
	\end{align}
	where  $c_{i,j}=\mathbbm{1}\left\{\frac{\vec{X}_{i,j}\cdot \vec{u}}{\|\vec{X}_{i,j}\|\|\vec{u}\|} \geqq  \frac{e}{r}\right\}$ and
	\begin{align}
		G_{i,j}&=\begin{cases}
			g_tg_r&\text{if }\frac{\vec{X}_{i,j}\cdot \vec{u}}{\|\vec{X}_{i,j}\|\|\vec{u}\|} \geqq  \frac{r^2+e^2-d_g^2}{2 r e },\\
			g_r&\text{if }\frac{\vec{X}_{i,j}\cdot \vec{u}}{\|\vec{X}_{i,j}\|\|\vec{u}\|} <  \frac{r^2+e^2-d_g^2}{2 r e }.
		\end{cases}
	\end{align}
\end{theorem}

\begin{IEEEproof}
Since all satellites visible to the typical user contribute to the total received signal power, the total interference is $T = \sum_{i,j: \text{visible}}		pG_{i,j}H\|X_{i,j}-u\|^{-\alpha},$
where $G_{i,j}=G(\|X_{i,j}-u\|).$
Leveraging the inner product of $\vec{X}_{i,j}$ and $\vec{u},$ the set of $i,j$ visible to the typical user is 	\begin{align}
	\{i,j:\text{visible}\}
	&\!= \left\{ i\in N_o,j\in N_s\left| \frac{\vec{X}_{i,j}\cdot \vec{u}}{\|\vec{X}_{i,j}\|\|\vec{u}\|} \geqq  \frac{e}{r}\right.\right\}\label{eq:ij:visible},
\end{align}
where $\vec{X}_{i,j}$ is given by Eqs. \eqref{vecx}, \eqref{vecy}, and \eqref{vecz}. Finally, let $c_{i,j}=\mathbbm{1}\left\{\frac{\vec{X}_{i,j}\cdot \vec{u}}{\|\vec{X}_{i,j}\|\|\vec{u}\|} \geqq  \frac{e}{r}\right\}$. We have
\begin{align}
	\cL_T(s)
	&= \bE_{\bar{\theta},\bar{\omega}}[\bE\left[\exp(-sT)\right]]\nnb\\
	&=\bE_{\bar{\theta},\bar{\omega}}\left[\prod_{i,j: \textnormal{visible}}\cL_H\left(\frac{spG_{i,j}}{\|X_{i,j}-u\|^{\alpha}}\right)\right]\nnb\\
	&= \bE_{\bar{\theta},\bar{\omega}}\left[ e^{\left(\sum {c_{i,j}}\log\left(\cL_H\left(\frac{s		p G_{i,j}}{\|X_{i,j}-u\|^{\alpha}}\right)\right) \right)}\right],\label{eq:LT}
\end{align}
where we used that $H$ is a random variable with Laplace transform $\cL_H(s)$. In Eq. \eqref{eq:LT}, $G_{i,j} $ is either $g_t g_r$ or $g_r$ determined by the distance between $X_{i,j} $ and $u$. Exploiting the inner product of two vectors $\vec{X}_{i,j}$ and $\vec{u}$, we get
\begin{align}
	G_{i,j}&=\begin{cases}
		g_tg_r&\text{if }\frac{\vec{X}_{i,j}\cdot \vec{u}}{\|\vec{X}_{i,j}\|\|\vec{u}\|} \geqq  \frac{r^2+e^2-d_g^2}{2 r e },\\
		g_r&\text{if }\frac{\vec{X}_{i,j}\cdot \vec{u}}{\|\vec{X}_{i,j}\|\|\vec{u}\|} <  \frac{r^2+e^2-d_g^2}{2 r e }.
	\end{cases}\label{eq:gij}
\end{align}
Finally, applying the distribution of $\bar{\theta}$ and $\bar{\omega}$ yields the Laplace transform of the total interference of the typical user.
 \end{IEEEproof}

\begin{figure}
 	\centering
 	\includegraphics[width=.9\linewidth]{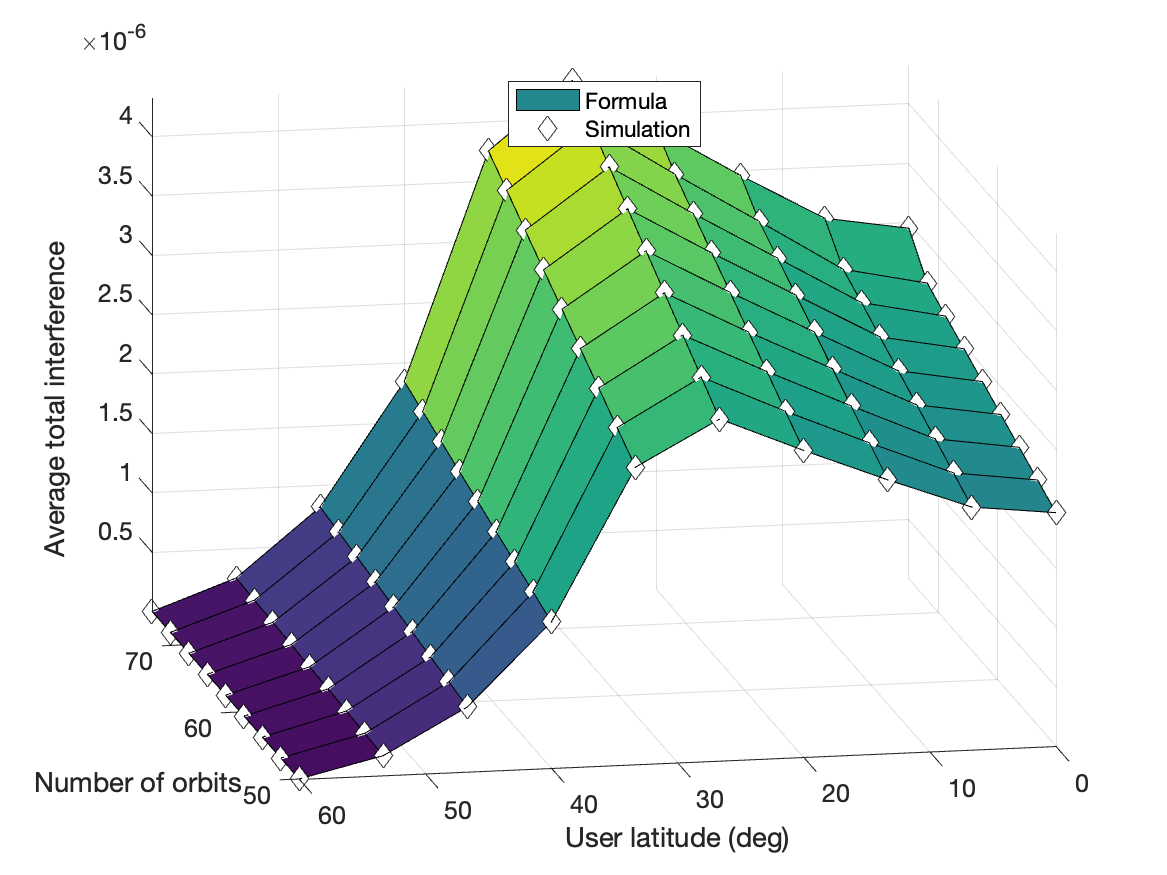}
 	\caption{The mean total interference. We use $\phi=33\degree,$ $N_s=75$, $B_w=10$ MHz, $p=20$ dBW, $g_t=24$ dBi, $g_r=1$, and $\eta = 5\degree$ (or $d_g=946$ km). }
 	\label{fig:theorem2}
 \end{figure}

Note that one can obtain the mean total interference by taking the expectation of $T$ w.r.t. $H$, $\bar{\theta}$, and $\bar{\omega}$ as follows:  \begin{align}
	\bE[T]=&k\int_{0}^{\frac{2\pi}{N_o}}\int_{0}^{\frac{2\pi}{N_s}} 	e^{\sum\left({c_{i,j}}\log\left(\frac{p\bE[H]	G_{{i,j}}}{\|X_{i,j}-u\|^{\alpha}}\right)\right)}\diff \bar{\omega} \diff \bar{\theta}.
\end{align}
Fig. \ref{fig:theorem2} shows the mean of the total interference. The simulation results confirm the accuracy of the derived formula. Note, the average total interference is the maximal for users at latitude $30\degree$.


\begin{theorem}\label{Theorem:5}
	The coverage probability of the typical user is
	\begin{equation}
	\bP(\SINR>\tau)=k\int_{0}^{\frac{2\pi}{N_s}}\int_0^{\frac{2\pi}{N_s}}\bE_{H_{1},...,H_n}
	\left[ \tilde{J} \right]\diff \bar{\omega} \diff \bar{\theta} ,\label{eq:theorem3}
	\end{equation}
where the constant $n$ is given by $|\left\{i,j: \text{visible}\right\}|-1$, $X_\star =\argmax \frac{\vec{X}_{i,j}\cdot \vec{u}}{\|\vec{X}_{i,j}\|\|\vec{u}\|},$ $G_\star = G(\|X_\star-u\|),$ and \begin{align}
	&\tilde{c}_{i,j}\! =\! \mathbbm{1}\!\left\{\frac{\vec{X}_{i,j}\cdot \vec{u}}{\|\vec{X}_{i,j}\|\|\vec{u}\|} \geqq  \frac{e}{r},\! \frac{\vec{X}_{i,j}\cdot \vec{u}}{\|\vec{X}_{i,j}\|\|\vec{u}\|}\neq \max\!\frac{\vec{X}_{i,j}\cdot \vec{u}}{\|\vec{X}_{i,j}\|\|\vec{u}\|}\! \!\right\},\nnb\\
	&\tilde{J}
	=\overbar{F}_H\left(\frac{\tau \|x_\star-u\|^{\alpha}}{pG_\star}\left(\sigma^2+\sum\!\!\frac{ph\tilde{c}_{i,j}G_{i,j}}{\|x_{i,j}-u\|^{\alpha}}\right)\right). \nnb
\end{align}
\end{theorem}

\begin{figure}
	\centering
	\includegraphics[width=.91\linewidth]{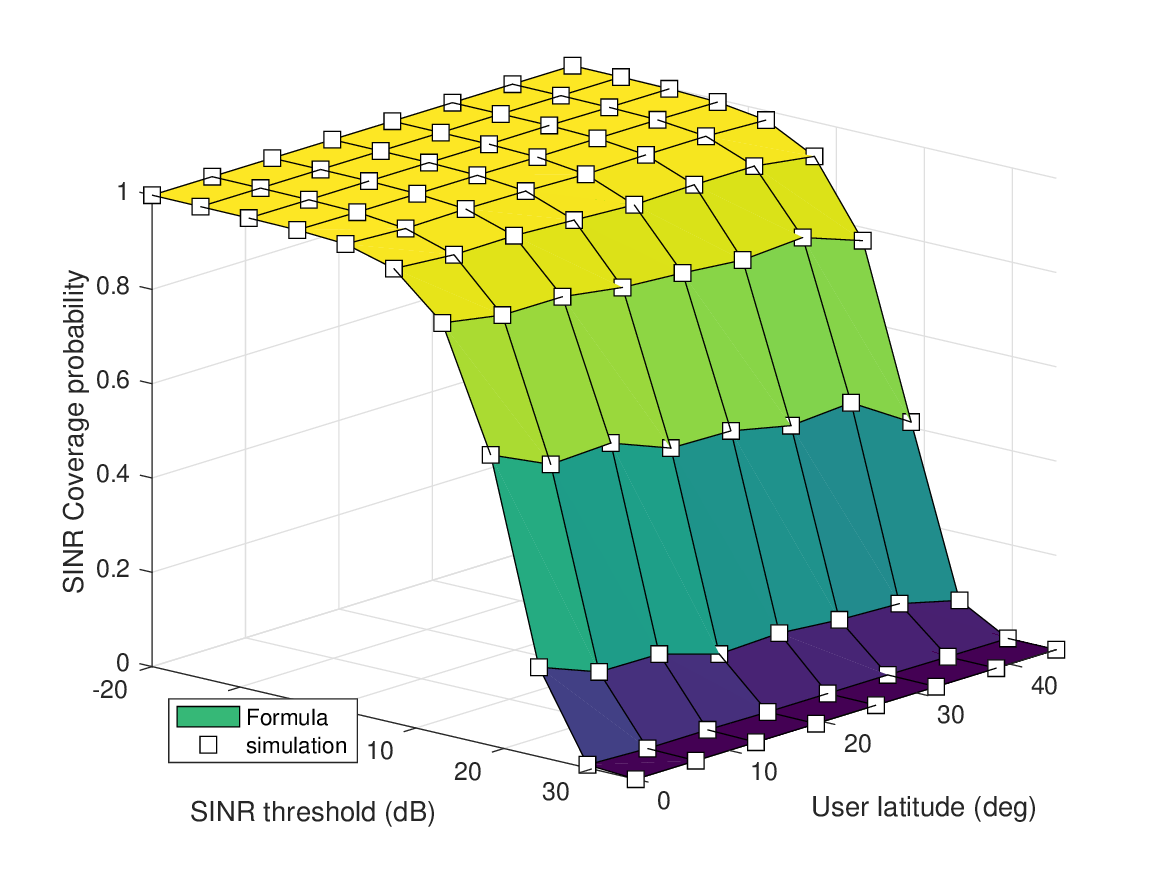}
	\caption{We use $\phi=33\degree,$ $N_o=25$, $N_s=30$, $g_t=30$ dB, $\eta=5\degree$, $g_r=1$, $T=300$ K, noise figure $N_f=7$ dB, $ B_w=10$ MHz, and $p=20$ dBW}
	\label{fig:theorem3}
\end{figure}

\begin{IEEEproof}
Since the typical user is assumed to be associated with its nearest satellite, the coverage probability is given by
\begin{align}
	\bP\left(\frac{pG_{\star}H_{\star}\|X_\star-u\|^{-\alpha}}{\sum\limits_{X_{i,j}\in\Psi(\bS_u)\setminus X_\star} p G_{i,j}H_{i,j} \|X_{i,j}\|^{-\alpha}+n}\right)\nnb,
\end{align}
where $G_\star = G(\|X_\star-u\|)$,  $G_{i,j} = G(\|X_{i,j}-u\|)$, and $X_\star =\argmax \frac{\vec{X}_{i,j}\cdot \vec{u}}{\|\vec{X}_{i,j}\|\|\vec{u}\|}.$

Let $l_\star$  be the orbit containing the nearest satellite and $\phi_\star$ be the satellite point process on it. Conditionally on $ X_\star,$ the interfering satellites are $\left\{X_{i,j} \in\Psi \setminus X_\star\left. \text{s.t.} \frac{\vec{X}_{i,j}\cdot \vec{u}}{\|\vec{X}_{i,j}\|\|\vec{u}\|} \geqq  \frac{e}{r} \right.\right\} $.

Now, the coverage probability of the typical user is
\begin{align}
	\bP(\SINR>\tau) =  \bP\left(\frac{pG_{\star}H_{\star}\|X_\star-u\|^{-\alpha}}{I+\sigma^2}>\tau \right),\nnb
\end{align}
where $I=\sum_{i,j:\text{interfering} } p G_{i,j}H_{i,j} {\|X_{i,j}-u\|}^{-\alpha}.$ Using the CDF of $H$, the coverage probability is given by the conditional expectation w.r.t. $\bar{\theta}$ and $\bar{\omega}$ as follows:
\begin{align}
	&\bP(\SINR>\tau)\nnb\\
	&=\bE_{\bar{\theta},\bar{\omega}} \left[\bP\left(\left.H> \frac{\tau \|x_\star-u\|^{\alpha} (I+\sigma^2)}{pG_{\star}}\right|\bar{\theta},\bar{\omega} \right)\right].\nnb
\end{align}
Let
$n$ be the number of elements in the set $\left\{i,j: \text{visible}\right\}$ minus one. Then, conditionally on $\bar{\theta}$ and $\bar{\omega},$ the random variable $I$ is a function of $n$ random variables. By conditioning further on the $n$ i.i.d. random variables  $\{H_k\}_{k=1,...,n}, $ we have
\begin{align}
	&\bP(\SINR>\tau)\nnb\\
	&=\bE_{\bar{\theta},\bar{\omega}}\left[\bE_{H_{1},\ldots,H_{n}}\left[\bP\left(H>\frac{\tau \|x_\star-u\|^{\alpha} (I+\sigma^2)}{pG_{\star}}\right)\right]\right]\nnb\\
	&	=\bE_{\bar{\theta},\bar{\omega}}\left[\bE_{H_{1},\ldots,H_{n}} \left[\tilde{J}\right]\right],
\end{align}
with
$\tilde{c}_{i,j}\! =\! \mathbbm{1}\!\left\{\frac{\vec{X}_{i,j}\cdot \vec{u}}{\|\vec{X}_{i,j}\|\|\vec{u}\|} \geqq  \frac{e}{r}, \frac{\vec{X}_{i,j}\cdot \vec{u}}{\|\vec{X}_{i,j}\|\|\vec{u}\|}\neq \max\left(\frac{\vec{X}_{i,j}\cdot \vec{u}}{\|\vec{X}_{i,j}\|\|\vec{u}\|}\right) \!\right\} $,
\begin{align}
	\tilde{J}
	&=\overbar{F}_H\left(\frac{\tau \|x_\star-u\|^{\alpha}}{pG_\star}\left(\sigma^2+\!\!\sum\!\!\frac{ph \tilde{c}_{i,j}G_{i,j}}{\|x_{i,j}-u\|^{\alpha}}\right)\right),\label{eq:sun}
\end{align}
where Eqs. \eqref{eq:ij:visible}  and \eqref{eq:gij} give the results for Eq. \eqref{eq:sun}.
\end{IEEEproof}
As in Theorem \ref{Theorem:2}, Theorem \ref{Theorem:5} yields a simpler expression for a Rayleigh fading scenario where $\overbar{F}_H(x)=e^{-x}$. Fig. \ref{fig:theorem3} illustrates the coverage probability for the Rayleigh fading. The figure confirms the accuracy of the derived formula.

The practical implication of employing the proposed geometric model lies in the fact that the derived formula computes the coverage probability of the typical user at a given latitude and it is equal to the coverage probability of all users at the same latitude. In addition, in the ergodic case of Theorem \ref{Theorem:periodic}, the expression theoretically matches the empirically-obtained time average of the coverage probability of any user at the given latitude, which otherwise requires an extensive amount of snapshots to evaluate using simulations.

\section{Conclusion}
In this paper, we established a stochastic geometry model for the Walker satellite constellation. We identified time invariance and proved other structural properties of the dynamical system, such as its periodicity and ergodicity. We then derived the typical performance of downlink communications as a function of key parameters, including the number of orbits, the number of satellites, the orbit inclination, and the user latitude. The derived metrics were interpreted based on the properties obtained from the dynamical system analysis. The proposed framework not only provides a tractable way to understand the interplay between network variables but also offers an intuitive method for designing and optimizing LEO satellite networks. Note the network performance with nonzero phase difference can be analyzed using similar techniques.

The established framework can be used to explore various applications based on multi-class Walker constellations. The framework could be used to analyze data harvesting via satellite relays or to investigate Internet routing.



\bibliographystyle{IEEEtran}
\bibliography{ref2}

\appendices

\section{No Satellite Collision}

\begin{lemma}
Consider the Walker constellation with parameters $\phi\ne 0$,
$N_0$, and $N_s$.
\begin{itemize}
\item Consider first antipodal pairs of orbits
(if the number of orbits $N_o$ is even, then there are $N_o/2$ such pairs;
if it is odd, then there are none).
If the number of satellites $N_s$ is even as well,
then the two antipodal satellites of any two antipodal orbits collide.
If $N_s$ is odd, then there is no collision on antipodal orbits.
\item
Consider now non-antipodal pairs of orbits. Each such pair is characterized
by an integer $k$, with $0 < k <\frac{N_o} 2$,
where $\frac{k2\pi}{N_o}$ is the angle (on the equator) between the two
descending and ascending points of the orbits of such a pair.
Then, a pair of satellites on these two orbital planes collide if and only if
$\tan\left(\frac{l\pi}{N_s}\right)=\cos(\phi)\tan\left(\frac{k\pi}{N_o}\right)$ for some integer $0< l < \frac{N_s} 2$.
\end{itemize}
\end{lemma}

\begin{IEEEproof} Orbits are numbered from 0 to $N_0-1$.
The orbit $0$ is the reference orbit. Let $0< k\le N_o-1$ be another orbit.
Note that if there are no collisions between orbit $0$ and
orbit $0 < k\le \frac{N_o} 2$,
then there are no collisions between orbit $\frac{N_o} 2 < k \le N_o -1$
and orbit $0$, by the rotational invariance.
Hence it is enough to check collisions between orbit $0$ and orbit $k$,
with $0 < k\le \frac{N_o} 2$.

Let $A_k$ (resp $B_k$) be the ascending (resp. descending) point
of orbit $k$, with $0 < k\le \frac{N_o} 2$.

We first consider the case where $N_0$ is even and $k=N_o/2$.
In this case, the two orbits are antipodal and they cross the equator
at the same points. Assume that at some given time, there
is a satellite of the orbit $0$ at $A_0$. Then, either $N_s$ is
even too and there is then a satellite of orbit $k$ at $B_k$.
Since $A_0=B_k$, there is hence a collision between satellites
of orbit $0$ and of orbit $N_0/2$, regardless of the inclination $\phi$.
If $N_s$ is odd, there is then no collision on these two orbits.
This proves the first statement.

Consider now two orbits $0$ and $k$, with $0 < k < \frac{N_o} 2$
(this is equilalent to all $0 < k \le \frac{N_o} 2$ if $N_o$ is even and
all $0 < k \le \frac{N_o} 2$ other than the antipodal orbits otherwise).
These two orbits meet at two points. Let $C_k$ be the
intersection point on the northern hemisphere.
Consider the spherical triangle $\triangle A_kC_kB_0$,
Then, the longitudinal difference between $A_k$ and $B_0$ is given by
$c=\pi-\frac{2\pi}{N_o}k$.

Let $a $ be the length of the arc $\arc{A_kC_k}$. Then, based on the
definition of the inclination, we have
$\sphericalangle C_kA_kB_0 = \sphericalangle C_kB_0A_k  =\phi $, namely
$\triangle A_kC_kB_0$ is a isosceles spherical triangle. Hence
the length of the arc $\arc{C_kB_0}$ is equivalent to $a$. Moreover,  since $a<\pi$, we can use Napier's analogy \cite{havil2010gamma}
for a isosceles spherical triangle $\alpha=\beta=\phi$ to get
\begin{align}\label{eq1}
	\cot(a)&=\cos(\phi)\frac{1+\cos(c)}{\sin(c)}\nnb\\
	&=\cos(\phi)\cot(c/2)\nnb\\
	&=\cos(\phi)\cot\left(\frac{\pi}{2} -\frac{k\pi}{N_o}\right)\nnb\\
	&=\cos(\phi)\tan\left(\frac{k\pi}{N_o}\right).
\end{align}
Then, there is a collision at $C_k$ between satellites of these two orbits
if and only if there is an integer $l$ such that
\begin{equation}
	\pi -2 a = \frac{2\pi}{N_s}l,\quad \mbox{mod } 2\pi.
\end{equation}
Indeed, for collision to occur, one must have two satellite
$S_0$ and $S_k $ of the orbits $0$ and orbit $k$, respectively,
at the intersection $C_k$ at the same time. Consider the time when the satellite $S_k$ is at $A_k,$ which is of an angular distance $a$ from the intersection point $C_k$ on the orbit k. Then, for a collision to occur at the same time, the satellite $S_0$ must be at the angular distance $2a$ from the point $B_0 $ on the orbit $0.$ In other words, there is a collision if and only if there is a satellite at angular distance $\pi-2a$ from the point $A_0$ on orbit $0$, namely the angle $\pi-2a $ is a multiple of $2\pi/N_s$.

As a result, for $0<a<\pi,$ there is a collision if and only
if for some $l$,
\begin{equation}
	\label{eq2}
	2a = \pi -\frac{l2\pi}{N_s}\quad \mbox{mod } 2\pi
\end{equation}
or equivalenly,
\begin{equation}
	a = \frac \pi 2 -\frac{l\pi}{N_s}\quad \mbox{mod } \pi.
\end{equation}
Since $0<a<\pi$, we have $-\frac{N_s} 2 < l < \frac{N_s} 2 $.

Hence, using Eq. \eqref{eq1}, we conclude that there is a collision
if and only if there exsists an integer $k$ with $0 < k < \frac{N_o} 2$
and an integer $l$ with $-\frac{N_s} 2 < l < \frac{N_s} 2 $ such that
$$\tan\left(\frac{l\pi}{N_s}\right) =\cos(\phi)\tan\left(\frac{k\pi}{N_o}\right),$$
Note that since $\cos(\phi)\ge 0$, if such an $l$ exists, it must be positive.
\end{IEEEproof}

A direct practical implication of the above lemma is as follows:
for all integers $N_s$ and $N_o$, if one of them is odd,
one can then always find an orbit inclination $\phi^\star$ \emph{arbitrarily}
close to a given $\phi $, and such that there is no collision of
satellites at any time in the proposed Walker constellation model.

\end{document}